\documentclass[twocolumn,prb,aps,showpacs]{revtex4}
\usepackage{dcolumn,amsfonts,bm}
\usepackage{graphicx, amsmath, amssymb, times}
\usepackage{setspace}
\begin{document}
\title{Pure Fulde-Ferrell-Larkin-Ovchinnikov state in optical lattices}
\author{A-Hai Chen}
\affiliation{Department of Physics, Zhejiang Normal University, Jinhua, Zhejiang Province, 321004, China}
\author{Gao Xianlong}
\email{gaoxl@zjnu.edu.cn}
\affiliation{Department of Physics, Zhejiang Normal University, Jinhua, Zhejiang Province, 321004, China}

\date{\today}

\begin{abstract}
We study the phase diagram of one-dimensional, two-component attractive fermions on optical lattices with an off-diagonal confinement. We identify in this system a pure Fulde-Ferrell-Larkin-Ovchinnikov (FFLO) state with spatially modulated superfluidity and polarization in a large parameter window, which provides an easier playground to detect the FFLO state experimentally. The clear signature of FFLO state is analyzed in the ground state properties, the pair correlation function, and the magnetic structure factor, using the density-matrix renormalization group method.
\end{abstract}
\maketitle

\textit{Introduction.} ---Experimental developments in cold atomic gases provide a promising testbed to realize the exotic quantum phases predicted in
condensed matter systems (for details see Ref. [1] for a recent review). Among these phases, there is an important state first predicted by Fulde, Ferrell, Larkin, and Ovchinnikov (FFLO) in 1964 for superconductors in high magnetic fields, characterized by cooper pairs with nonzero center-of-mass momentum and a spatially inhomogeneous order parameter.~\cite{FFLO}

Though the FFLO state has not been explicitly confirmed experimentally, its realization in ultracold atomic systems is becoming promising due to the great adjustability for almost all of the system parameters, for example, the spin imbalance realized by the magnetic fields in solid state physics can be easily simulated using the mismatched different particle species by radio-frequency fields.~\cite{RFF} While the FFLO state is difficult to detect in three-dimensional ultracold atomic systems due to the very small parameter window in free space,~\cite{Sheehy} the possibility of direct realization in the one-dimensional (1D) case increases due to the major part of the phase diagram being of the character of the FFLO correlation, based on analytical,~\cite{FFLO-theory,FFLO-ana} numerical,~\cite{FFLO-num} and experimental~\cite{FFLO exp} research.

Experimentally~\cite{FFLO exp} the density profiles of a spin imbalanced mixture of ultracold ${}^6$Li atoms trapped in an array of 1D tubes are measured with the standard absorption imaging. The density profiles detected meet the criteria of the FFLO state predicted analytically.~\cite{FFLO-ana} In a 1D system of harmonic confinement, a partially polarized core is surrounded by wings of either a fully paired or a fully polarized Fermi gas depending on the controlled imbalanced population in its two spin states. The partially polarized core in the density profile is proved of FFLO characteristics by calculating the pair momentum.~\cite{FFLO-ana} But in order to demonstrate directly that this pair state is of FFLO type, one needs to precisely measure the physical quantities peculiar to FFLO states, for example, the condensed fraction using the time-of-flight (TOF) measurement and the quantities other than density profile revealing the finite momentum of the pairing order parameter, like the correlations in the shot noise of fermion absorption images in TOF.~\cite{RFF,short noise,noise correlation} In a 1D optical lattice under the presence of a harmonic confining potential, Korolyuk et al. proposed that the change of the double occupancy due to the lattice depth modulation provides a clear evidence.~\cite{Korolyuk} The collective modes produced by the response of the ground state to time-dependent potentials give another possible signature of the FFLO phase.~\cite{Edge,Liu} Very recently, Kajala et al. proved that the TOF expansion manifests a direct signature of the FFLO state in the 1D system in which the expansion velocity of the unpaired particles matches the expected FFLO momentum.~\cite{Kajala}

In cold atom experiments, the necessary confining potential, which is usually a single-particle potential called diagonal confinement, leads to the emergence of coexisting states~\cite{Rigol} and inevitably makes the detected signals complicated. In the repulsive case the diagonal confinement leads to the absence of true incompressible Mott insulating phases and a real superfluid-to-Mott transition.~\cite{ODC} Though Scarola et al. pointed out that edge effects in a trapped Fermi-Hubbard system can be filtered out with a direct probe independent of inhomogeneity~\cite{Scarola} by relating the core compressibility to changes in the double occupancy, people hope to realize directly a pure Mott insulating or other exotic phases in the trapped system by making use of state-of-the-art cold atomic techniques. In order to achieve such a goal, an off-diagonal confinement method was pointed out by Rousseau et al.,~\cite{ODC,ODC2} where site-dependent hopping integrals are designed with amplitude decreasing while moving away from the center of the lattice and vanishing towards the boundaries.

In this paper, we show that for the system in the off-diagonal confinement of population imbalance, a pure FFLO state is realized at a large range of system parameters, and thus, in this case, the common multi-phase structure in the diagonal confinement system is avoided, which facilitates the detection of such an elusive state in the cold atomic system. The clear signature of FFLO state is analyzed through calculating the pair correlation function and the magnetic structure factor, using density-matrix renormalization group (DMRG) method.

\textit{The model and phase diagram.} ---We consider fermions confined to a 1D optical lattice with $L$ sites and lattice constant $a=1$. The Hamiltonian is given by
\begin{equation}\label{1}
H=-\sum_{\langle i,j\rangle,\sigma}t_{ij}(c_{i,\sigma}^{\dag}c_{j,\sigma}+h.c.)+
U\sum_{i=0}^{L-1} n_{i\uparrow}n_{i\downarrow}\,,
\end{equation}
where $c^{(+)}_{i,\sigma}$ is a fermionic annihilation (creation) operator acting on site $i$ with pseudospin $\sigma$ representing the hyperfine states and the sum over $\langle i,j\rangle$ is restricted to the nearest neighbours. The spatially varying tunneling is of inverted parabola designed as
\begin{equation}\label{1a}
t_{ij}=t(i+j+1)(2L-i-j-1)/L^2,
\end{equation}
which vanish at the edges avoiding the escape of the particles from the edges, namely, $t_{-1,0}=t_{L-1,L}=0$. The properties of the FFLO state in a 1D optical lattice of mass imbalance have been studied intensively,~\cite{tsigma,Orso,Gu} where $t_{ij}$ in Eq. (1) is replaced by $t_\sigma$. They found that the typical FFLO modulation is preserved in the gapped phase but not the quasi-long-range order~\cite{Orso} and if the system is partially polarized, the ground-state FFLO state is more stable in the presence of mass imbalance.~\cite{Gu}

We study the case of attractive on-site interactions ($U<0$) with population imbalance. In the following, we take $u=U/t$ with $t$ as the units of energy. The number of fermions of each species is $N_\sigma=\sum_i \langle c^+_{i,\sigma} c_{i,\sigma}\rangle$ with $N=N_\uparrow+N_\downarrow$ and the spin $\sigma=\uparrow, \downarrow$. The filling factor, the magnetization, and the polarization are $n=N/L$, $S_z=(N_\uparrow-N_\downarrow)/2$, and $p=(N_\uparrow-N_\downarrow)/N$, respectively. We work in the canonical ensemble with fixed total number of particles. In the following we choose a typical experimentally accessible optical lattice of $L=70$ sites and assume $N_\uparrow > N_\downarrow$ without loss of generality.

Experimentally, this off-diagonal confinement system can be realized by a holographic method with a mask under the present-day ultracold atomic techniques.~\cite{ODC} Technically, the off-diagonal confinement system is somehow equivalent to the smooth boundary conditions introduced by Veki\'{c} and White aiming at reducing finite-size effects and extrapolating to the thermodynamic limit on relatively small systems,~\cite{White} and is similar to the sine-squared or sinusoidal deformation applied to 1D quantum Hamiltonians.~\cite{Gendiar} In our study we discuss a partially polarized system of inverted parabola tunneling.~\cite{difftij}
\begin{figure} [htb]
\centerline{\scalebox{0.4}
{\includegraphics{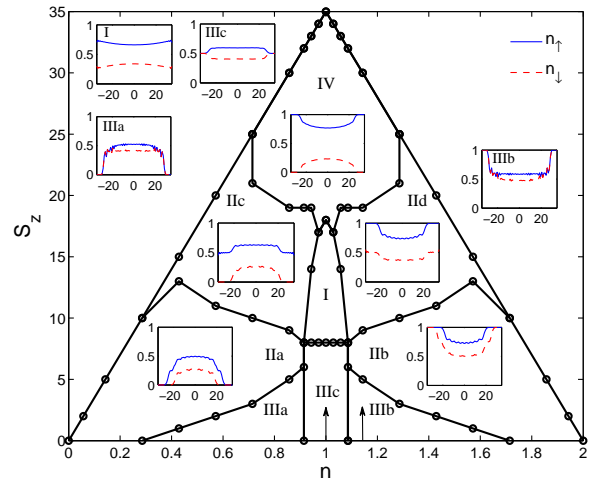}}}
\caption{(Color online) The ground-state phase diagram of the off-diagonal confinement system in the $S_z$-$n$ plane for $u=-4$. The pure FFLO superfluid phase (I) locates in the parameter range of $8<S_z<18$ and $0.93<n<1.1$. Other composite phases we labeled are FFLO-FP (IIa, IIc), FFLO-FI-FS (IIb), FFLO-FIa (IId), FFLO-BCS (IIIa, IIIc), FFLO-BCS-FS (IIIb), and FFLO-FIb (IV), respectively (details see text). The density distributions for spin-up and spin-down particles in the different regions are shown versus $i$ for typical situations as insets. The arrows indicate the direction of the different phases while increasing the magnetization $S_z$ for fixed fermions of $N=70$ and $80$.}
\label{fig:one}
\end{figure}
\begin{figure} [htb]
\begin{center}
\includegraphics*[width=0.9\linewidth]{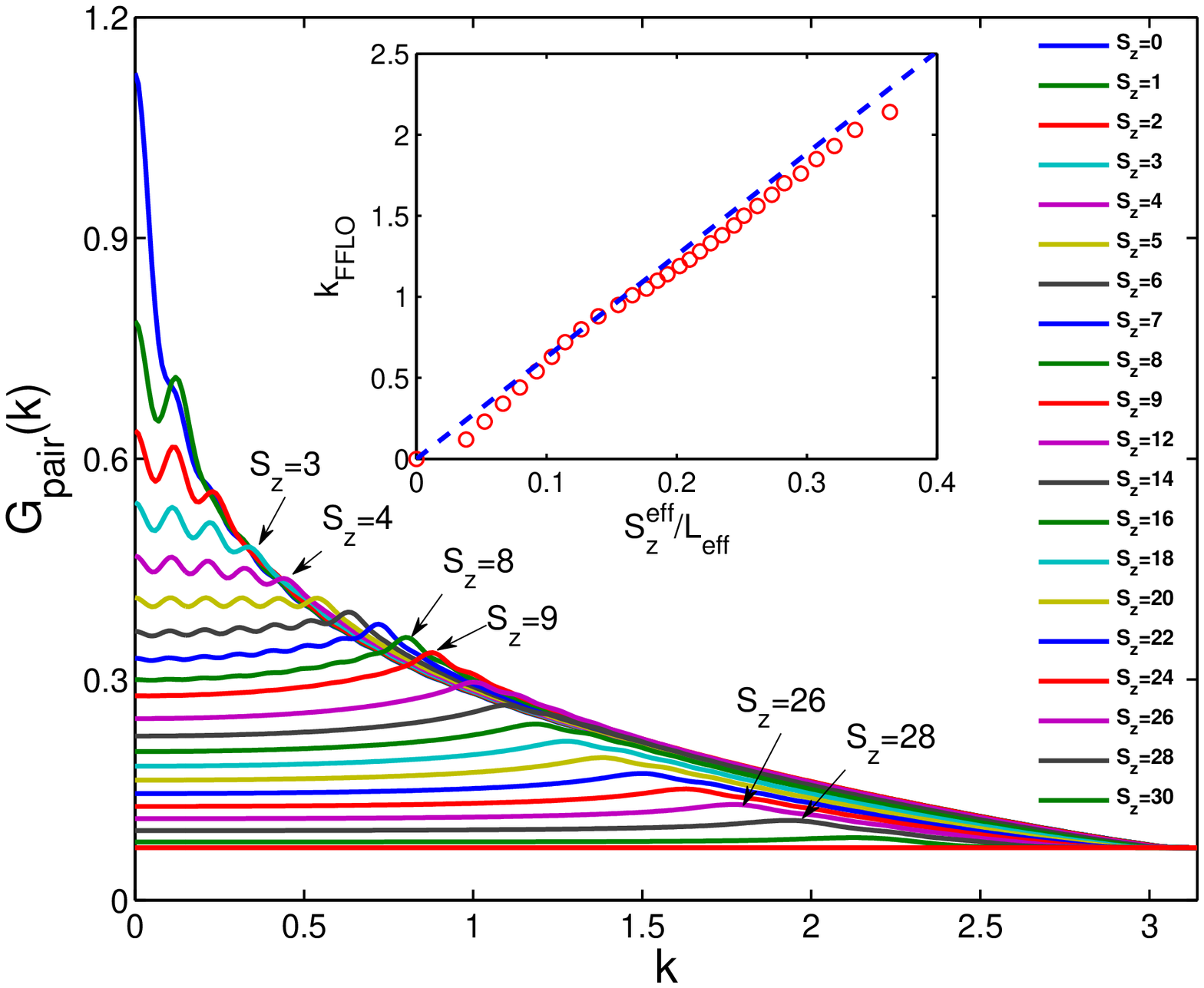}
\includegraphics*[width=0.9\linewidth]{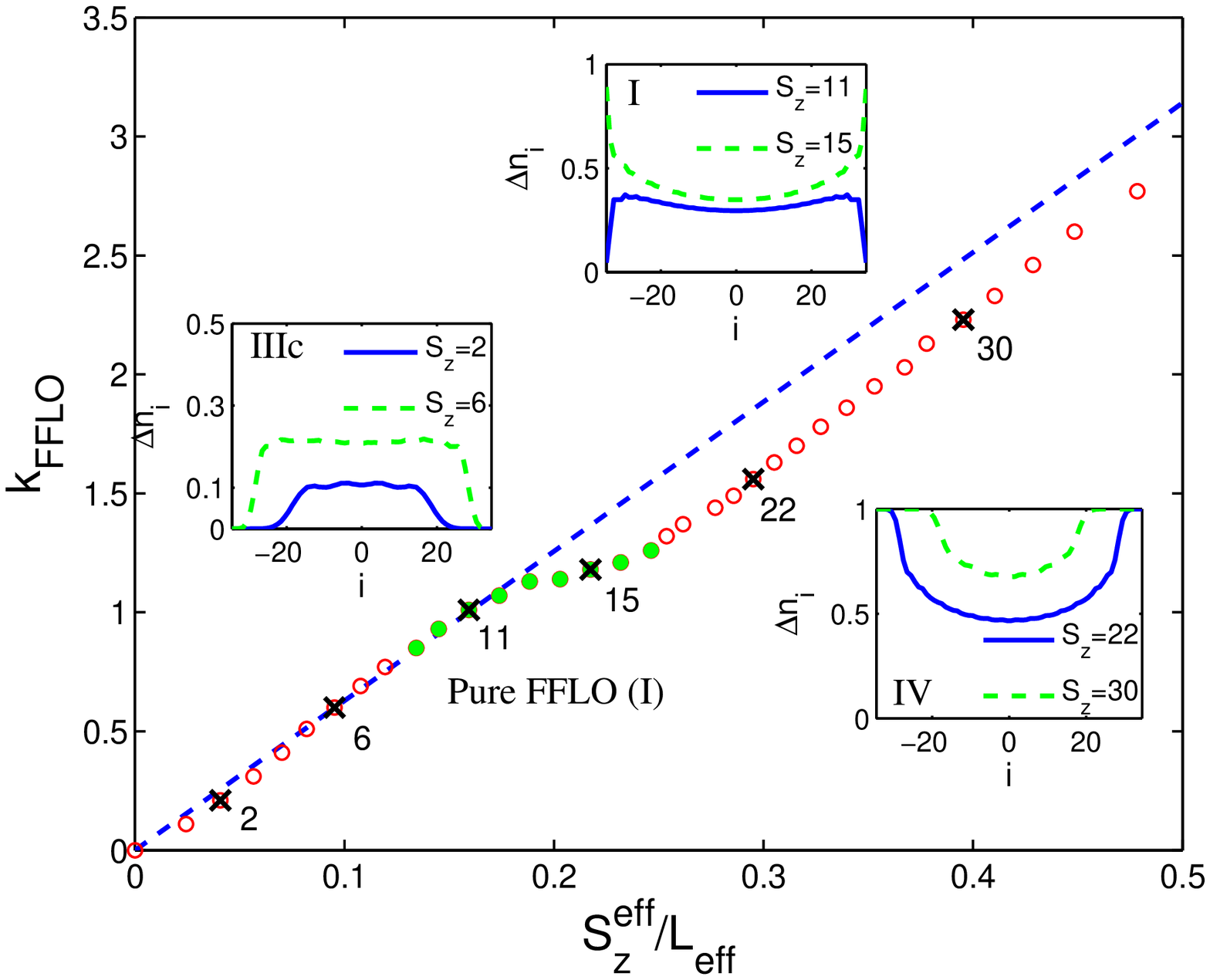}
\caption{(Color online) Upper panel: the pair momentum distributions for the magnetization varying from $S_z=0$ to $30$ (from top to bottom) in a system of $N=80$ fermions. The inset shows that the momentum of the FFLO state $k_{\rm FFLO}$, at which the $G_{\rm pair}$ is peaked, as a function of the effective magnetization at the core $S^{\rm eff}_z/L_{\rm eff}$. With the increase of $S_z$, the phases of the system change from $\rm IIIb\rightarrow IIb\rightarrow IId\rightarrow IV$ for $N=80$ and from $\rm IIIc\rightarrow I\rightarrow IV$ for $N=70$ (in the lower panel), which is indicated in Fig.~\ref{fig:one} by arrows. The dashed line represents $k_{\rm FFLO}=\Delta \tilde{k}_{\rm F}$. Lower panel: the position of the FFLO peak as a function of $S^{\rm eff}_z/L_{\rm eff}$ in a system of $N=70$ fermions. $k_{\rm FFLO}$ in the pure FFLO region is denoted by the full circle. In the inset, the profiles of the populations difference $\Delta n_i$ are shown for two different magnetizations $S_z$ in the corresponding phase region. The crosses with $S_z$ values are pertinent to the insets. See the text for a discussion of the relations between the linearity of $k_{\rm FFLO}$ and $\Delta n_i$ in different phases. The interaction strength used here is $u=-4$.}
\label{fig:two}
\end{center}
\end{figure}
\begin{figure}
\centerline{\scalebox{0.4}
{\includegraphics{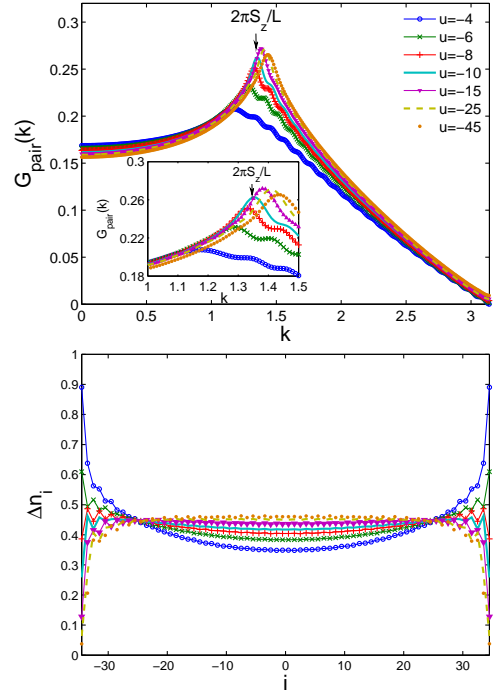}}}\caption{
(Color online) Upper panel: the pair momentum distribution for different interaction strength. The parameters used here are $S_z=15$ and $N=70$. The one in the lower-left magnifies the region around the position of the peaks.
Lower panel: the corresponding density difference $\Delta n_i=n_{i\uparrow}-n_{i\downarrow}$ in the real space. The notations are the same as those in upper panel.}
\label{fig:three}
\end{figure}
\begin{figure}
\centerline{\scalebox{0.4}
{\includegraphics{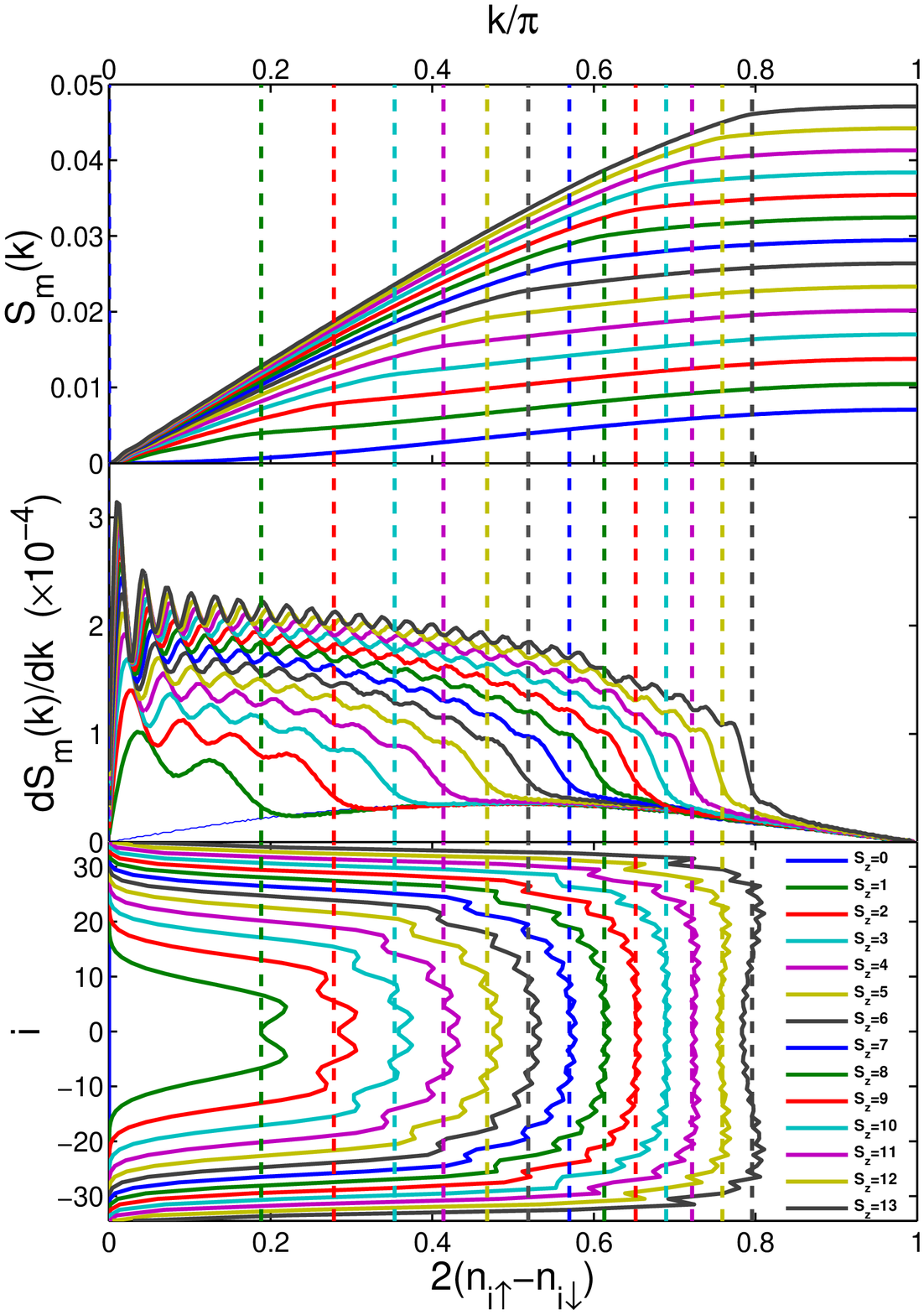}}}\caption{
(Color online) Upper panel: magnetic structure factor as a function of $k/\pi$ with fixed number of fermions N=70 and interaction strength $u=-10$ for different magnetization $S_z$ (from top $S_z=0$ to bottom $S_z=30$). Middle panel: the corresponding momentum derivative of the magnetic structure factor. Lower panel: corresponding density imbalance distributions, $2(n_{i\uparrow}-n_{i\downarrow})$. The dashed lines mark the matching positions of the kink in the structure factor, the corresponding jump in the derivative of the magnetic structure factor, and the central density imbalance.}
\label{fig:four}
\end{figure}

The ground-state phase diagram is shown in Fig.~\ref{fig:one}. The typical density distributions in different ranges are displayed as insets. We find that a partially polarized phase ($0<n_{i\uparrow}, n_{i\downarrow}<1$) exists in the core of the system at finite polarization. Among all the different phases, we identify an important pure FFLO superfluid phase which is partially polarized in the whole space of the system in part I, for $S_z\in [8, 18]$ and $n\in [0.93, 1.1]$. We would like to point out that in the diagonal confinement system the pure FFLO state happens only at a certain critical polarization. Other phases identified are, a fully polarized phase ($0<n_{i\uparrow}<1, n_{i\downarrow}=0$, FP), an unpolarized BCS phase ($n_{i\uparrow}=n_{i\downarrow}\ne 0$, BCS), a pure Fock state of localized pairs, namely a band-insulating state, at high densities ($n_{i\uparrow}= n_{i\downarrow}=1$, FS), and fully polarized ferromagnetic insulating phase for one species ($n_{i\uparrow}=1$) and partially filled ($n_{i\downarrow} \ne 0$, FIa) or empty ($n_{i\downarrow}=0$, FIb) for another. All the regions in Fig.~\ref{fig:one} except part I are composed of coexisting phases, either of two-phase structure: FFLO-FP (IIa, IIc), FFLO-FIa (IId), and FFLO-BCS (IIIa, IIIc), FFLO-FIb (IV), or of three-phase structure: FFLO-FI-FS (IIb) and FFLO-BCS-FS (IIIb). The relation between the parts of IIa and IIb, IIc and IId, and IIIa and IIIb, can be understood by the particle-hole symmetry, $c_{i\sigma} \rightarrow (-1)^i c^+_{i\sigma}$, mapping the Hamiltonian (1) onto itself. As a result, if the particle number in part IIa and part IIb satisfies $N^a_\uparrow+N^b_\downarrow=N^a_\downarrow+N^b_\uparrow$ and $N^a+N^b=2L$, the densities in the corresponding parts are related by $n^a_{i\sigma}=1-n^b_{i\bar{\sigma}}$. Here $\bar{\sigma}=-\sigma$.

\textit{Pairing correlations.} ---Let us now come to the discussion of the pairing properties. We first compute the pair correlation function, which is defined as,
\begin{equation}\label{PCF}
G(i,j)=\langle \Delta_i^{\dag}\Delta_j\rangle
\end{equation}
with $\Delta_i=c_{i\uparrow} c_{i\downarrow}$. Its Fourier transform is given by
\begin{equation}\label{PCFa}
G_{\rm pair}(k)=\frac{1}{2L}\sum_{i,j}G (i,j)e^{ik(i-j)}\,.
\end{equation}
Feiguin and Heidrich-Meisner confirmed that for the attractive Hubbard model in a harmonic confinement the momentum pair distribution has peaks at the mismatch of the Fermi surfaces.~\cite{Feiguin} This quantity is displayed in Fig.~\ref{fig:two} for varying magnetization from $S_z=0$ to $30$.

For the unpolarized case, the ground state is a BCS state characterized by a sharp peak centered at momentum $k=0$ in the pair momentum distribution $ G_{\rm pair}(k)$. For the population imbalanced Fermi system of two spin species, two different Fermi surfaces exist and the ground state is a 1D FFLO state with a finite center-of-mass momentum ($k \ne 0$), which can be understood as an order parameter of the system. In the homogeneous system, the momentum of the FFLO state $k_{\rm FFLO}$, at which the $G_{\rm pair}$ is peaked, is, $k_{\rm FFLO}=k_{\rm F \uparrow}-k_{\rm F \downarrow}=\pi(n_{\uparrow}-n_{\downarrow})=2\pi S_z/L=\pi n p \neq 0$ with $k_{\rm F \sigma}=\pi n_{\sigma}=\pi N_{\sigma}/L$. In the inhomogeneous system, we define $\Delta \tilde{k}_{\rm F}=2\pi S^{\rm eff}_z/L_{\rm eff}$, where $S^{\rm eff}_z$ is obtained by integrating $\langle S^i_z \rangle$ over the effective partially polarized FFLO region $L_{\rm eff}$ with $S^i_z$ defined as $S^i_z=(n_{i\uparrow}-n_{i\downarrow})/2$.~\cite{Feiguin} We notice that, in Fig.~\ref{fig:two}, the value of $k_{\rm FFLO}$ increases with $S_z$ as predicted by the FFLO picture. But in the inhomogeneous system under the off-diagonal confinement, where the hopping amplitude $t_{ij}$ is spatially varying, the relation between $k_{\rm FFLO}$ and $\Delta \tilde{k}_{\rm F}$ is not as simple as that under the diagonal confinement. For $N=80$, in the cases where the density differences in the bulk become inhomogeneous, $k_{\rm FFLO}$ deviates slightly from $\Delta \tilde{k}_{\rm F}$. For the system of $N=70$ (lower panel), the phases of the system undergoes $\rm IIIc\rightarrow I\rightarrow IV$ while increasing $S_z$. Starting from $S_z=13$, $k_{\rm FFLO}$ greatly deviates from $\Delta \tilde{k}_{\rm F}$ at which $\Delta n_i=n_{i\uparrow}-n_{i\downarrow}$ in the bulk becomes largely inhomogeneous. The FFLO peak around $k_{\rm FFLO}$ comes from the FFLO state in the bulk of the density, while the central peaks (of oscillating character) come from the paired BCS wings.

For the pure FFLO states, we show in Fig.~\ref{fig:three} the effects of the interactions on the pair momentum distributions. We find the optimal interaction strength ($u\approx -15$) where the peak of $G_{\rm pair}$ reaches maximum. The position of the peak matches exactly the linearity relation of $k_{\rm FFLO}=2\pi S_z/L\approx 1.346$ at $u\approx -10$, where $\Delta n_i$ is almost homogeneous, in the whole range of the system except at the edges.

\textit{Magnetic structure factor.} ---We further reveal fingerprints of the FFLO state by analyzing the magnetic structure factor (MSF), which is the Fourier transform of spin-spin correlations,
\begin{eqnarray}\label{MSF}
S_m(k)=\frac{1}{2L}\sum_{i,j}(\langle S^i_z S^j_z\rangle -\langle S^i_z\rangle \langle S^j_z\rangle)e^{ik(i-j)}\,.
\end{eqnarray}
Spin-spin correlations can be detected in a non-destructive way via spatially resolved quantum polarization spectroscopy. Roscilde et al. showed analytically and numerically that $2k_{\rm FFLO}$-modulation of spin-spin correlations in the MSF is a direct fingerprint of pairing in an imbalance system and serves as a good quantity in detecting the FFLO state.~\cite{Roscilde}

We display the MSF in Fig.~\ref{fig:four} for $N=70$ and $u=-10$ with different polarization values. Comparing to the unpolarized case, there is an obvious kink in the MSF with finite $S_z$. To have a clear signature of this kink, we show the momentum derivative of the MSF which exhibits a marked jump, where the density imbalance distributions in the bulk, $\Delta n_i$, reveal a constant structure. The match of the location of the kink with the bulk region of the density deteriorates when the density difference becomes inhomogeneous, like the cases in the lower panel of Fig.~\ref{fig:two} when $S_z>13$. We notice that the density difference in the lower panel is obviously oscillatory for small polarizations. The wavelength of the oscillations is given by $\lambda=2\pi/|k_{\rm F\uparrow}-k_{\rm F\downarrow}|$, which gives a confirmation that the FFLO state is inhomogeneous and exhibits real-space modulations.~\cite{Wolaka}

{\it Conclusions.} ---We have considered a 1D optical lattice of off-diagonal confinement, modeled by a single-band Hubbard model with spatially varying hopping amplitudes. We showed that with population imbalance the dominant pairing mechanism in the ground state is FFLO characterized by a finite center-of-mass momentum. We predicted based on the $S_z$-$n$ phase diagram that a pure FFLO state exists at a large parameter range, greatly different from the diagonal confinement system where only a critical polarization can achieve that. We further revealed that the partially polarized densities are the results of the finite center-of-mass momentum pairing through calculating the pair correlation function and the spin-spin correlation. The flexibility in ultracold atomic experiment could be used in designing the spatially varying hopping amplitude, and thus, provides an ideal playground for realizing such a FFLO superfluid state. Furthermore, the existence of the pure FFLO phase could facilitate an unambiguous experimental observation of FFLO pairing in atomic systems free of the influence from other quantum phases at edges. The MSF which can be detected by the quantum polarization spectroscopy signal also provides a direct signature of finite center-of-mass momentum FFLO pairing.

Acknowledgements. We acknowledge discussions with F. Gleisberg. This work is supported by the National Natural Science Foundation of China under Grant No.~11174253, and Zhejiang Provincial Natural Science Foundation of China under Grant No. R6110175. The DMRG simulations were performed using the ALPS libraries.~\cite{DMRG}


\begin{thebibliography}{99}
\bibitem{review}
        M. Lewenstein, A. Sanpera, V. Ahufinger, B. Damski, A. Sen (De), and U. Sen, Adv. Phys. {\bf 56}, 243 (2007);
        I. Bloch, J. Dalibard, and W. Zwerger, Rev. Mod. Phys. {\bf 80}, 885 (2008);
        S. Giorgini, L. P. Pitaevskii, and S. Stringari, Rev. Mod. Phys. {\bf 80}, 1215 (2008).
\bibitem{FFLO}
        P. Fulde and R. A. Ferrell, Phys. Rev. {\bf 135}, A550 (1964);
        A. I. Larkin and Yu. N. Ovchinnikov, Sov. Phys. JETP {\bf 20}, 762 (1965).
\bibitem{RFF}
        M. Greiner, C. A. Regal, J. T. Stewart, and D. S. Jin, Phys. Rev. Lett. {\bf 94}, 110401 (2005).
\bibitem{Sheehy}
        D. E. Sheehy and L. Radzihovsky, Phys. Rev. Lett. {\bf 96}, 060401 (2006);
        K. Machida, T. Mizushima, and M. Ichioka, Phys. Rev. Lett. {\bf 97}, 120407 (2006);
        J. Kinnunen, L. M. Jensen, and P. T\"{o}rm\"{a}, Phys. Rev. Lett. {\bf 96}, 110403 (2006).
\bibitem{FFLO-theory}
        K. Yang, Phys. Rev. B {\bf 63}, 140511(R) (2001);
        Phys. Rev. Lett. {\bf 95}, 218903 (2005);
        E. Zhao and W. V. Liu, Phys. Rev. A {\bf 78}, 063605 (2008);
        X. W. Guan, M. T. Batchelor, C. Lee, and M. Bortz, Phys. Rev. B {\bf 76}, 085120 (2007);
        J. Y. Lee and X. W. Guan, Nucl. Phys. B {\bf 853}, 125 (2011).
\bibitem{FFLO-ana}
        G.~Orso, Phys. Rev. Lett. {\bf 98}, 070402 (2007);
        H.~Hu, X.-J.~Liu, and P. D.~Drummond, Phys. Rev. Lett. {\bf 98}, 070403 (2007).
\bibitem{FFLO-num}
        M. Casula, D. M. Ceperley, and E. J. Mueller, Phys. Rev. A {\bf 78}, 033607 (2008);
        M. Rizzi, M. Polini, M. A. Cazalilla, M. R. Bakhtiari, M. P. Tosi, and R. Fazio, Phys. Rev. B {\bf 77}, 245105 (2008);
        G. G. Batrouni, M. H. Huntley, V. G. Rousseau, and R. T. Scalettar, Phys. Rev. Lett. {\bf 100}, 116405 (2008);
        M. Tezuka and M. Ueda, Phys. Rev. Lett. {\bf 100}, 110403 (2008);
        G. Xianlong and R. Reza, Phys. Rev. A {\bf 77}, 033604 (2008).
\bibitem{FFLO exp}
        Y. Liao, A. S. C. Rittner, T. Paprotta, W. Li, G. B. Partridge, R. G. Hulet, S. K. Baur, and E. J. Mueller, Nature {\bf 467}, 567 (2010).
\bibitem{short noise}
        E. Altman, E. Demler, and M. D. Lukin, Phys. Rev. A {\bf 70}, 013603 (2004).
\bibitem{noise correlation}
        A.~L\"{u}scher, R. M.~Noack, and A. M.~L\"{a}uchli, Phys. Rev. A {\bf 78}, 013637 (2008).
\bibitem{Korolyuk}
        A. Korolyuk, F. Massel, and P. T\"{o}rm\"{a}, Phys. Rev. Lett. {\bf 104}, 236402 (2010).
\bibitem{Liu}
        X.-J. Liu, H. Hu, and P. D. Drummond, Phys. Rev. A {\bf 76}, 043605 (2007).
\bibitem{Edge}
        J. M. Edge and N. R. Cooper, Phys. Rev. Lett. {\bf 103}, 065301 (2009).
\bibitem{Kajala}
        J. Kajala, F. Massel, and P. T\"{o}rm\"{a}, Phys. Rev. A {\bf 84}, 041601(R) (2011) .
\bibitem{Rigol}
        M.~Rigol, A.~Muramatsu, G. G.~Batrouni, and R. T.~Scalettar, Phys. Rev. Lett. {\bf 91}, 130403 (2003).
\bibitem{ODC}
        V. G. Rousseau, G. G. Batrouni, D. E. Sheehy, J. Moreno, and M. Jarrell, Phys. Rev. Lett. {\bf 104}, 167201 (2010);
        V. G. Rousseau, K. Hettiarachchilage, M. Jarrell, J. Moreno, and D. E. Sheehy, Phys. Rev. A {\bf 82}, 063631 (2010).
\bibitem{Scarola}
        V. W. Scarola, L. Pollet, J. Oitmaa, and M. Troyer, Phys. Rev. Lett. {\bf 102}, 135302 (2009).
\bibitem{ODC2}
        J. Carrasquilla and F. Becca, Phys. Rev. A {\bf 82}, 053609 (2010).
\bibitem{Orso}
        G. Orso, E. Burovski, and T. Jolicoeur, Phys. Rev. Lett. {\bf 104}, 065301 (2010).
\bibitem{tsigma}
        M. A. Cazalilla, A. F. Ho, and Th. Giamarchi, Phys. Rev. Lett. {\bf 95}, 226402 (2005);
        B. Wang, H.-D. Chen, and S. Das Sarma, Phys. Rev. A {\bf 79}, 051604 (R) (2009);
        G. G. Batrouni, M. J. Wolak, F. H\'{e}bert, and V. G. Rousseau, EPL {\bf 86}, 47006 (2009);
        E. Burovski, G. Orso, and T. Jolicoeur, Phys. Rev. Lett. {\bf 103}, 215301 (2009);
        G. Roux, E. Burovski, and T. Jolicoeur, Phys. Rev. A {\bf 83}, 053618 (2011).
\bibitem{Gu}
        W.-L. Lu, Z.-G. Wang, S.-J. Gu, and H.-Q. Lin, arXiv:0902.1021v1.
\bibitem{White}
        M. Veki\'{c} and S. R. White, Phys. Rev. Lett. {\bf 71}, 4283 (1993);
        Phys. Rev. B {\bf 53}, 14552 (1996).
\bibitem{Gendiar}
        A. Gendiar, M. Daniska, Y. Lee, and T. Nishino, Phys. Rev. A {\bf 83}, 052118 (2011), and references therein.
\bibitem{difftij}
        For curiosity, different spatially dependent hoppings are tested, with the one of a sine-squre deformation, $t_i=\sin^2 (i\pi/L)$ studied by Gendiar et al.~\cite{Gendiar} and the other with $t_i=C_{M-i}$ for $1\le i\le M$; $1$ for $M<i\le L-M$; and $C_{i-L+M}$ for $L-M < i <L$, where, $C_m=\{[1-\tanh [(m-M/2)/(m(1-m/M))] \}/2$ studied by Veki\'{c} and S. R. White~\cite{White}. We found that, the density behaves in the similar way in the bulk. For the pair correlation functions, there are differences in the position and the height of the peak. All of these designs in the hopping amplitude are designated as reducing the boundary influence. We prefer to use the off-diagnol confinement in optical lattices to study the pure FFLO state while the latter two are more difficult to realize in experiments than the parabolic one used in this paper.
\bibitem{Feiguin}
        A. E. Feiguin and F. Heidrich-Meisner, Phys. Rev. B {\bf 76}, 220508(R) (2007);
        Phys. Rev. Lett. {\bf 102}, 076403 (2009);
        F. Heidrich-Meisner, A. E. Feiguin, U. Schollw\"{o}ck, and W. Zwerger, Phys. Rev. A {\bf 81}, 023629 (2010).
\bibitem{Roscilde}
        T. Roscilde, M. Rodr\'{i}guez, K. Eckert, O. Romero-Isart, M. Lewenstein, E. Polzik, and A. Sanpera, New J. Phys. {\bf 11}, 055041 (2009).
\bibitem{Wolaka}
        M. J. Wolaka, V. G. Rousseau, and G. G. Batrouni, Comput. Phys. Commun. {\bf 182}, 2021 (2010).
\bibitem{DMRG}
        A. F. Albuquerque et al. (ALPS Collaboration), J. Magn. Magn. Mater. {\bf 310}, 1187 (2007).
\end{thebibliography}
\end{document}